\documentclass[submission,copyright]{eptcs}

\providecommand{\event}{Wivace 2013}

\usepackage{breakurl}
\usepackage{amsmath}
\usepackage{graphicx}
\usepackage{subfig}
\usepackage{sidecap}

\usepackage{amssymb}
\usepackage{todonotes}

\title{Analysis of the spatial and dynamical properties \\ of a multiscale model of intestinal crypts}

\author{Giulio Caravagna $\quad$  Alex Graudenzi $\quad$ Marco Antoniotti
\institute{Universit\`{a} degli Studi di Milano-Bicocca, \\Dipartimento di Informatica, Sistemistica e Comunicazione\\ Viale Sarca 336, 20126 Milano, Italy}
\email{\{marco.antoniotti, giulio.caravagna, alex.graudenzi\}@disco.unimib.it} \bigskip\\
Giovanni De Matteis
\institute{Department of Mathematics and Information Sciences,\\
Northumbria University, \\Newcastle Upon Tyne,
NE2 1XE, UK.}
\email{giovanni.dematteis@northumbria.ac.uk}
}

\def\titlerunning{Investigating intestinal crypt homeostasis  through multiscale modeling}
\def\authorrunning{Giulio Caravagna {\em et al.}}
\begin{document}
\maketitle

\section*{Extended Abstract}
\begin{figure}[!h]
\begin{center}
\includegraphics[width=0.85\textwidth]{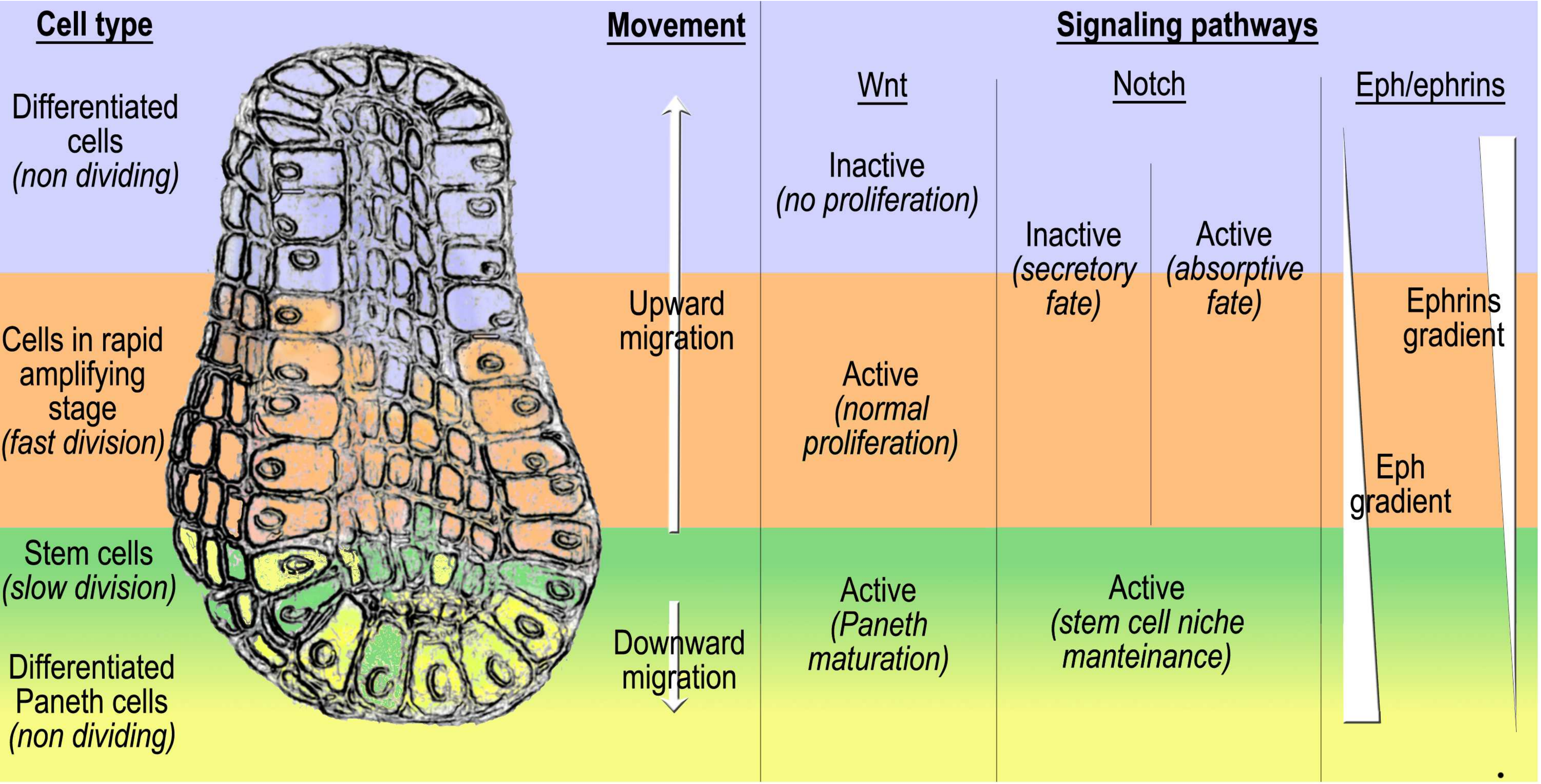}
\caption{{\bf Schematic crypt representation.} 
Crypt morphology, migration directions and the involved signaling pathways (Wnt, the Notch and the Eph/ephrins). Four types of cells are represented, according to Figure \ref{fig:tree}. All cells but stem and Paneth migrate upward. Picture taken from \cite{DeMatteis2013}.
 }
\label{fig:crypt}
\end{center}
\end{figure}

Intestinal crypts are invaginations in the connective tissue of the human intestine and are supposed to be the site in which tumors originate, starting from the alteration of certain genes and pathways \cite{Alberts2007}. 
Considering that, so far, only a partial characterization of the key features of intestinal crypts has been accomplished, the in-depth investigation of their structural and dynamical properties is of crucial importance, with particular regard to the emergence and development of colorectal cancer (CRC) \cite{DeMatteis2013}.   The basic elements involved in the crypt dynamics and the morphology of the crypt are shown in Figure \ref{fig:crypt}.

\begin{SCfigure}
  \centering
  \includegraphics[width=0.45\textwidth]%
    {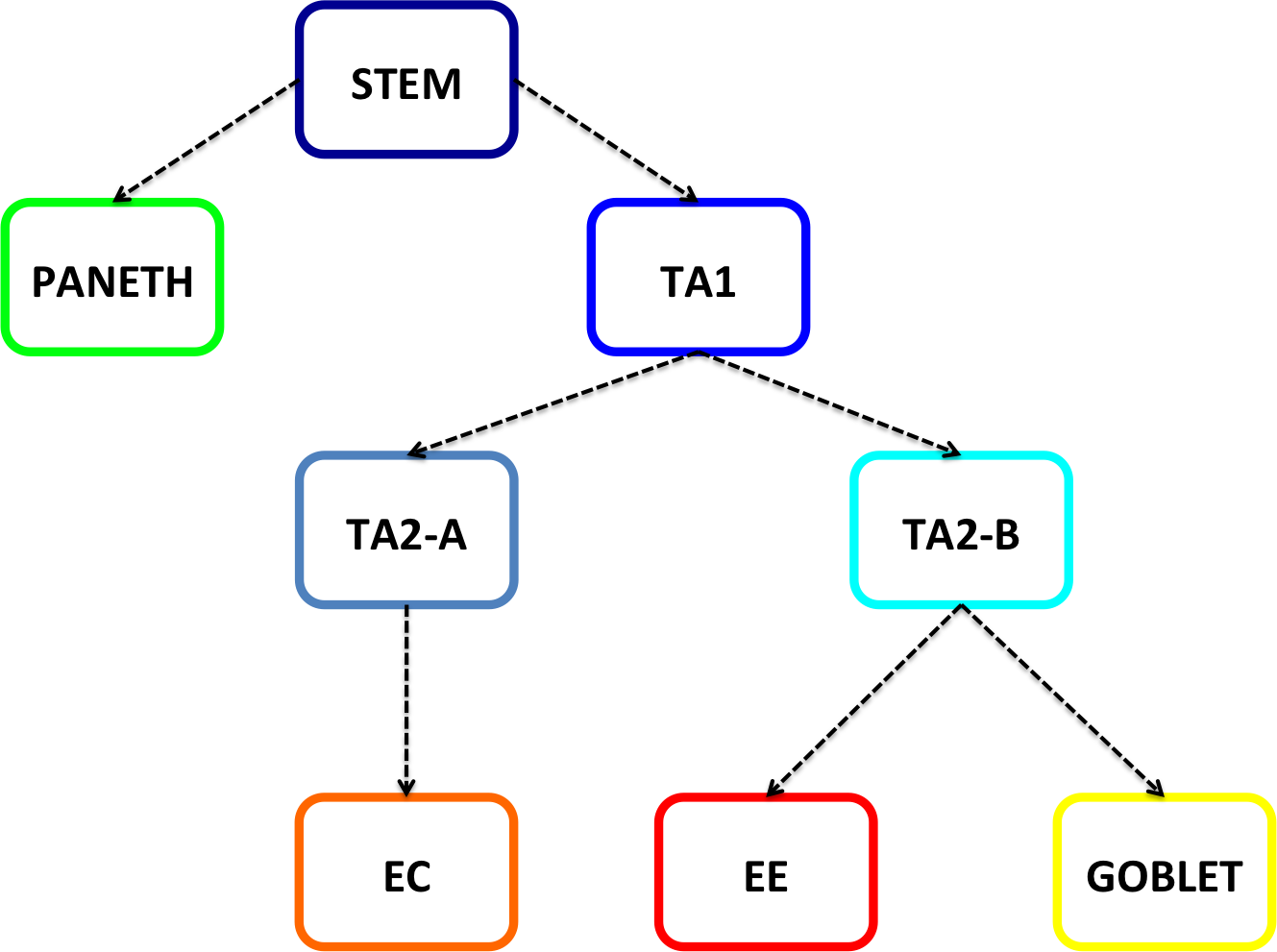}
\caption{{\bf Differentiation tree for crypts.} We consider stem cells, transit-amplifying stage cells (TA1, TA2-A, TA2-B) and fully differentiated cells (paneth, goblet cells, enterocytes and enteroendocrine). These represent  the principal cell types of the epithelium of the small intestine \cite{Alberts2007}. Cellular differentiation is modeled as a stochastic process whose dynamics emerge from a  {\em Noisy Random Boolean Network} representation  of a {\em Gene Regulatory Network}, along the lines of  \cite{serra2010}.}
\label{fig:tree}
\end{SCfigure}

Upon these premises, a multiscale model of intestinal crypt dynamics, which has already been conceived in its preliminary form \cite{GraudenziWivace2012}, is presented. 
The model combines a spatial/morphological {\em Cellular Potts Model} (CPM) \cite{GG1992} with a {\em Noisy Random Boolean Network} (NRBN) model of {\em Gene Regulatory Network} \cite{serra2010}.  
The CPM is a statistical mechanics model in which cells are represented as lattice sites in a 2D cellular automaton. The changes in cell shapes and positions are induced by the stochastic motion minimizing the overall system energy. In our model cells:
\begin{itemize}
\item[$(i)$] grow in size up to a limit value, then undergo mitosis;
\item[$(ii)$] are preferentially close to cells of the same type, according to the known {\em Differential Adhesion Hypothesis} \cite{Steinberg1962};
\item[$(iii)$]  differentiate according to a precise differentiation scheme.
\end{itemize}
 In our model  we consider 8 cellular populations: stem cells, transit-amplifying stage cells and fully differentiated cells, as shown in Figure \ref{fig:tree}.

\begin{SCfigure}
  \centering
  \includegraphics[width=0.7\textwidth]%
    {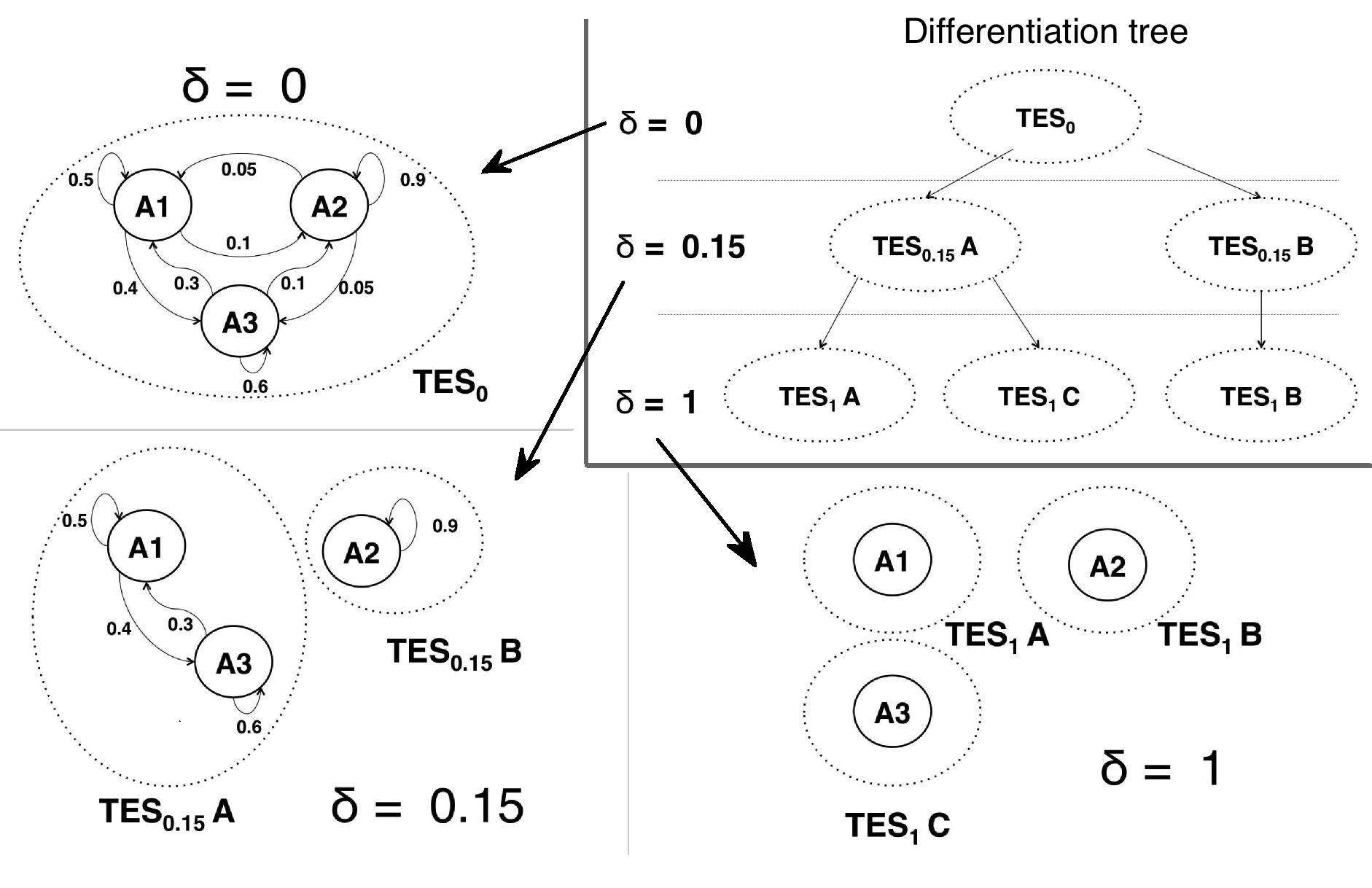}
\caption{{\bf NRBN model.} We show the differentiation tree emerging from a NRBN (not shown). A1, A2 and A3 are shortnames for the NRBN attractors, and the edges weights are the relative frequency of transitions among attractors as obtained by noise-induced perturbations. The values of $\delta$ represent the noise thresholds, according to \cite{Villani2011}.}
\label{fig:tes}
\end{SCfigure}

The length of the cell cycle, the size growth dynamics and the differentiation paths are driven by the emergent properties of the internal NRBN dynamics. 
In particular, a NRBN is a discrete synchronous dynamical system in which genes are represented as Boolean nodes that influence each other through interaction pathways, modeled as Boolean functions. Sets of system steady states (i.e. the attractors) represent gene activation patterns associated to biologically stable functions. The underlying hypothesis is that cell types present more or less stable behaviors in presence of stochastic noise. Thus, the degree of differentiation of any specific cell type is related to a noise-resistance threshold. In this modeling approach the less differentiated cells are associated to a lower resistance, and vice versa. This phenomenon is known as stochastic (i.e. noise-induced) differentiation \cite{Villani2011}.  We graphically represent it in Figure \ref{fig:tes}.

The key novelties of the model lay in: 
\begin{itemize}
\item[$(i)$] the focus on the emergent dynamical behavior of the GRNs driving the spatial dynamics;\item[$(ii)$] the possibility of simulating different gene-level perturbations, e.g. mutations and pathway alterations, with obvious reference to the emergence and development of cancer.\end{itemize}
The first simulations of the model helped to decipher the sufficient conditions to achieve the general homeostasis of the crypt, respecting cell populations proportion, cell sorting and coordinate migration.  
To this end, distinct crypts ruled by randomly generated NRBNs and with different initial configurations of the lattice were simulated.  

\paragraph{Results.} 
Simulations suggest that the stochastic differentiation process is itself sufficient to ensure homeostasis in the asymptotic states, where the proportion among cell populations is stable in time, independently of the initial conditions. 
Besides, it was shown that some key physical quantities, e.g. speed and size of cells, are in accordance with the self-sustaining of the system and with experimental data \cite{wong2010}.  
A correlation between the direction and velocity of the movement of the neighboring cells was quantitatively evaluated, hinting at the existence of a coordinate movement of close cells. 
Finally, quantitative measures of the level of spatial order, e.g. the MoranÕs index and the Pearson correlation coefficient, proved that the crypt achieves a sufficient level of order depending on the initial configuration, hence hinting at the deep importance of the morphogenesis processes.  

In general, it was shown that a complex dynamical phenomenon such as intestinal crypt homeostasis does not require any particular tuning or constraint regarding growth, division and differentiation, which can be efficiently ruled by the emerging dynamics of simplified models of GRNs.  

The undergoing research is focused on the simulation of gene-perturbations, e.g. knockouts, in order to investigate their effects on the crypt, with a particular focus on the potential disruption of homeostasis and/or emergence of aberrant spatial structures.  

\newpage
\nocite{*}
\bibliographystyle{eptcs}
\bibliography{biblio}

\end{document}